\documentclass{article}
\usepackage{amsfonts}
\begin{document}

\title{Correlation function of the velocity field of a thin suspended
liquid film}
\author{Doriano Brogioli}

\maketitle

\begin{abstract}
In this paper, I consider a thin suspended liquid film, surrounded by
a different fluid. Examples of such a system are soap films and
liquid crystal films, surrounded by air. They are considered good models
for two dimensional fluid dynamics, although air drag is known to introduce
strong deviations.
I exactly solve the hydrodynamic equations
of the system composed by the liquid layer and the surrounding fluid
in order to evaluate the correlation function of the thermally
excited velocity fluctuations. Both the temporal and the spatial 
statistical properties are the ones of a two dimensional, hypotetical
fluid, only for temporal and spatial frequencies higher than critical values;
at lower frequencies they show significative deviations due to the interactions
of the liquid layer with the surrounding fluid. 
\end{abstract}

\section{Introduction}

A thin, freely suspended liquid layer is generally considered a system 
that follows two dimensional hydrodynamics, on length scales longer
than the layer thickness. Many experiments have been performed on soap films,
that consist of two monolayers of amphiphilic molecules around a layer of
water. The layer thickness ranges from the $4\mathrm{nm}$
of the Newton black film to many micrometers; features down to a fraction
of a millimeter should be considered two dimensional
\cite{goldburg95_1}.

Also smectic A
liquid crystals can form films, by a different effect. It spontaneously
form a layered structure, with the elongated molecules alligned to the
surface normal. When a film is drawn from 
this kind of liquid crystal, it is formed by a set of monolayers, ranging from 
two to many hundreds \cite{bechhoefer97}; the thickness can be selected with
extreme accuracy. Because of the absence of translational order in the 
monolayers, the film behaves as a two dimensional system.

Many papers 
\cite{goldburg95_1,goldburg95_2,goldburg96,couder84,couder89}
describe experiments in which turbulent flows have been detected, and 
discuss the agreemen with two dimensional hydrodynamics. One result is that
the role of the air surrounding the film cannot be neglected. In turbulent 
flows, it damps the turbulent vortices. From a theoretical point of wiev, an 
exact evaluation of this effect is extremely difficult; for example,
Rivera et al. \cite{rivera00}
modeled this damping as a linear drag term in the two dimensional 
Navier-Stokes equations,
in order to describe velocity fluctuations observed in a turbulent flow.
They evaluated the amplitude of the drag term from experimental
data; from this value, they found the dissipation due to air friction to be a
significant energy dissipation mechanism in their experimental system.

Another well known effect of air friction on thin liquid films
can be found in diffusion of particles, whose diameter is about the thickness
of the layer. Neglecting air friction, the mobility of a particle in a
two dimensional system should be infinite, that is, a particle subjected to
a steady force should accelerate indefinitely. This would result in 
an infinite diffusion coefficient.
Saffman \cite{saffman76} found that air friction reduces the mobility to
a finite value: he evaluated the diffusion coefficient and found that it
strongly depends on the viscosity of the surrounding fluid, and diverges
as the viscosity vanishes. Saffman formula has been tested both in
soap films \cite{cheung96} and in liquid crystal films \cite{bechhoefer97}.

The diffusion coefficient can be evaluated once we know the 
correlation function of the thermally excited velocity fluctuations; this
means that the velocity correlation
function must be strongly affected by air friction. In this paper, I
evaluate the correlation function of the velocity field of a thin
suspended liquid film, by solving the hydrodynamic equation
of the whole system, including the liquid film and the fluid surrounding it.

Then I discuss the result, showing that both the temporal and the spatial 
statistical properties are the ones of a two dimensional, hypotetical
fluid, only for temporal and spatial frequencies higher than critical values;
at lower frequencies they show significative deviations due to the interactions
of the liquid layer with the surrounding fluid. 

Last, I derive approximately the value of diffusion
coefficient, and I compare it with Saffman formula.

\section{Derivation of the power spectrum of the velocity fluctuations}

I describe the film and the fluid surrounding it by using Landau's
fluctuating hydrodynamics equations:
\begin{equation}
\left\{
\begin{array}{l}
\rho_0\frac{\partial}{\partial t}\vec{u}_0\left(\vec{x},t\right)=
\eta_0\nabla^2\vec{u}_0\left(\vec{x},t\right)-\vec{\nabla}p_0
\left(\vec{x},t\right)+\frac{\eta_1}{h}\left[
\frac{\partial}{\partial z}\vec{u}_1\left(\vec{x},z=0,t\right)-
\frac{\partial}{\partial z}\vec{u}_2\left(\vec{x},z=0,t\right)
\right]
\\
\rho_1\frac{\partial}{\partial t}\vec{u}_{1,2}\left(\vec{x},z,t\right)=
\eta_1\nabla^2\vec{u}_{1,2}\left(\vec{x},z,t\right)-\vec{\nabla}p_{1,2}
\left(\vec{x},z,t\right)
\end{array}
\right.
\end{equation}
where $\vec{u}_1\left(\vec{x},z,t\right)$ and
$\vec{u}_2\left(\vec{x},z,t\right)$ are tree dimensional vector fields,
describing the velocities in the space above and below the film;
$\vec{u}_0\left(\vec{x},t\right)$ is a two dimensional vector field,
describing the velocity in the film; $\eta_0$ and
$\eta_1$ are the shear viscosity of the film and the surrounding fluid;
$\rho_0$ and $\rho_1$ are the densities of the film and the surrounding
fluid, $h$ is the thickness of the film;
$p_{1,2}\left(\vec{x},z,t\right)$ and $p_0\left(\vec{x},t\right)$ are
the pressures.

I suppose that both the fluids are
incompressible, the film is placed at $z=0$, and the fluid does
not slip at the surface of the sheet:
\begin{eqnarray}
\vec{\nabla} \cdot \vec{u}_0\left(\vec{x},t\right)&=&0
\label{eq_incomprimibilita1}\\
\vec{\nabla} \cdot \vec{u}_{1,2}\left(\vec{x},z,t\right)&=&0
\label{eq_incomprimibilita2}\\
\vec{u}_{1,2}\left(\vec{x},z=0,t\right)&=&
\left[\vec{u}_0\left(\vec{x},t\right),0\right]
\end{eqnarray}
The pressures $p_{1,2}\left(\vec{x},z,t\right)$ and
$p_0\left(\vec{x},t\right)$ must be determined in order to fulfill the
incompressibility equations Eqs (\ref{eq_incomprimibilita1},
\ref{eq_incomprimibilita2}).

In Fourier space:
\begin{equation}
\left\{
\begin{array}{l}
\left(-i\rho_0\omega+\eta_0q^2\right)\vec{u}_0\left(\vec{q},\omega\right)=
i\vec{q}p_0\left(\vec{q},\omega\right)-i\frac{\eta_1}{h}\int{
q_z\left[\vec{u}_1\left(\vec{q},q_z,\omega\right)-\vec{u}_2\left(\vec{q},q_z,\omega\right)\right]dq_z}\\
\left[-i\rho_1\omega+\eta_1\left(q^2+q_z^2\right)\right]
\vec{u}_{1,2}\left(\vec{q},q_z,\omega\right)=
i\left[\vec{q},q_z\right]p_{1,2}\left(\vec{q},q_z,\omega\right)
\end{array}
\right.
\end{equation}
with the conditions:
\begin{eqnarray}
\vec{q} \cdot \vec{u}_0\left(\vec{q},\omega\right)&=&0\\
\left[\vec{q},q_z\right] \cdot
\vec{u}_{1,2}\left(\vec{q},q_z,\omega\right)&=&0\\
\int{\vec{u}_{1,2}\left(\vec{q},q_z,\omega\right)dq_z}
&=&
\left[\vec{u}_0\left(\vec{q},\omega\right),0\right]
\end{eqnarray}
where the Fourier transform is defined as follows:
\begin{equation}
f\left(\vec{q}\right)=\frac{1}{\left(2\pi\right)^D}\int{e^
{\displaystyle i\vec{q}\cdot\vec{x}}f\left(\vec{x}\right)d\vec{x}}
\end{equation}

Now I use the incompressibility equations Eqs (\ref{eq_incomprimibilita1},
\ref{eq_incomprimibilita2}) in order to evaluate the pressures:
\begin{equation}
\left\{
\begin{array}{l}
\left(-i\rho_0\omega+\eta_0q^2\right)\vec{u}_0\left(\vec{q},\omega\right)=
-i\frac{\eta_1}{h}\left(\mathbb{I}-\vec{q}\frac{\vec{q}}{q^2}\right)\int{
q_z\left[\vec{u}_1\left(\vec{q},q_z,\omega\right)-\vec{u}_2\left(\vec{q},q_z,\omega\right)\right]dq_z}\\
\left[-i\rho_1\omega+\eta_1\left(q^2+q_z^2\right)\right]
\vec{u}_{1,2}\left(\vec{q},q_z,\omega\right)=0
\end{array}
\right.
\end{equation}
with the conditions:
\begin{eqnarray}
\vec{q} \cdot \vec{u}_0\left(\vec{q},\omega\right)&=&0\\
\left[\vec{q},q_z\right] \cdot
\vec{u}_{1,2}\left(\vec{q},q_z,\omega\right)&=&0\\
\int{\vec{u}_{1,2}\left(\vec{q},q_z,\omega\right)dq_z}
&=&
\left[\vec{u}_0\left(\vec{q},\omega\right),0\right]
\end{eqnarray}

Then, I express the vectors on a suitable basis:
\begin{eqnarray}
\vec{u}_0\left(\vec{q},\omega\right)&=&u_{0R}\left(\vec{q},\omega\right)\frac{\left[q_x,q_y\right]}{\sqrt{q_x^2+q_y^2}}+u_{0T}\left(\vec{q},\omega\right)\frac{\left[q_y,-q_x\right]}{\sqrt{q_x^2+q_y^2}}\\
\vec{u}_{1,2}\left(\vec{q},q_z,\omega\right)&=&
u_{1,2R}\left(\vec{q},q_z,\omega\right)\frac{\left[q_x,q_y,q_z\right]}{\sqrt{q_x^2+q_y^2+q_z^2}}+
u_{1,2T}\left(\vec{q},q_z,\omega\right)\frac{\left[q_y,-q_x,0\right]}{\sqrt{q_x^2+q_y^2}}+\\
&&
u_{1,2Z}\left(\vec{q},q_z,\omega\right)\frac{\left[-q_xq_z,-q_yq_z,q_x^2+q_y^2\right]}
{\sqrt{q_x^2+q_y^2}\sqrt{q_x^2+q_y^2+q_z^2}}
\end{eqnarray}
The subscripts $R$ and $T$ label the radial and tangential components of
both $\vec{u}_0$ and $\vec{u}_{1,2}$, and $Z$ labels the component
of $\vec{u}_{1,2}$ perpendicular to the film. The equations become:
\begin{equation}
\left\{
\begin{array}{l}
\left(-i\rho_0\omega+\eta_0q^2\right)u_{0T}\left(\vec{q},\omega\right)=
-i\frac{\eta_1}{h}\int{
q_z\left[u_{1T}\left(\vec{q},q_z,\omega\right)-u_{2T}\left(\vec{q},q_z,\omega\right)\right]dq_z}\\
\left[-i\rho_1\omega+\eta_1\left(q^2+q_z^2\right)\right]
u_{1,2T}\left(\vec{q},q_z,\omega\right)=0\\
\left[-i\rho_1\omega+\eta_1\left(q^2+q_z^2\right)\right]
u_{1,2Z}\left(\vec{q},q_z,\omega\right)=0\\
u_{0R}\left(\vec{q},\omega\right)=0\\
u_{1,2R}\left(\vec{q},q_z,\omega\right)=0\\
\int{u_{1,2T}\left(\vec{q},q_z,\omega\right)dq_z}
=u_{0T}\left(\vec{q},\omega\right)\\
\int{u_{1,2Z}\left(\vec{q},q_z,\omega\right)
\frac{1}{\left|\left[\vec{q},q_z\right]\right|}dq_z}=0\\
\int{u_{1,2Z}\left(\vec{q},q_z,\omega\right)
\frac{q_z}{\left|\left[\vec{q},q_z\right]\right|}dq_z}=0
\end{array}
\right.
\end{equation}

Only the components $u_{0T}$ and $u_{1,2T}$ are needed in order to
evaluate $u_0$; in the following, I will drop the subscript $T$.
\begin{equation}
\left\{
\begin{array}{l}
\left(-i\rho_0\omega+\eta_0q^2\right)u_{0}\left(\vec{q},\omega\right)=
-i\frac{\eta_1}{h}\int{
q_z\left[u_{1}\left(\vec{q},q_z,\omega\right)-u_{2}\left(\vec{q},q_z,\omega\right)\right]dq_z}\\
\left[-i\rho_1\omega+\eta_1\left(q^2+q_z^2\right)\right]
u_{1,2}\left(\vec{q},q_z,\omega\right)=0\\
\int{u_{1,2}\left(\vec{q},q_z,\omega\right)dq_z}
=u_{0}\left(\vec{q},\omega\right)
\end{array}
\right.
\end{equation}

In real space:
\begin{equation}
\left\{
\begin{array}{l}
\frac{\partial}{\partial t}u_{0}\left(\vec{q},t\right)=
-\nu_0q^2u_{0}\left(\vec{q},t\right)+ \frac{\eta_1}{h\rho_0}
\left[\frac{\partial}{\partial z}
u_{1}\left(\vec{q},z=0,t\right)-\frac{\partial}{\partial z}
u_{2}\left(\vec{q},z=0,t\right)\right]\\
\frac{\partial}{\partial t}
u_{1,2}\left(\vec{q},z,t\right)= -\nu_1q^2
u_{1,2}\left(\vec{q},z,t\right)+ \nu_1\frac{\partial^2}{\partial
z^2} u_{1,2}\left(\vec{q},z,t\right)\\
u_{1,2}\left(\vec{q},z=0,t\right) =u_{0}\left(\vec{q},t\right)
\end{array}
\right.
\end{equation}
where $\nu=\eta/\rho$.

In order to simplify the following calculations, I impose periodic boundary
conditions on $z$, with a period $L$; later, I will evaluate the limit
for $L \to \infty$. I express the hydrodynamic variables on a 
complete orthonormal basis:
\begin{eqnarray}
u_{1,2}\left(\vec{q},z,t\right)&=&
\sum_{n=0}^{\infty}{\left\{
a_n^{\vec{q}}\left(t\right)f^A_n\left(z\right)+
b_n^{\vec{q}}\left(t\right)f^B_n\left(z\right)
\right\}}+
c^{\vec{q}}\left(t\right)f^C\left(z\right)
\\
\label{eq_scomposizione_u0}
u_{0}\left(\vec{q},t\right)&=&
\sum_{n=0}^{\infty}{
b_n^{\vec{q}}\left(t\right)f^B_n\left(0\right)
}+
c^{\vec{q}}\left(t\right)f^C\left(0\right)
\end{eqnarray}
The eigenfunctions are:
\begin{eqnarray}
f_n^A\left(z\right)&=&N_n^A\sin\left[A_n\left(z-L/2\right)\right]\\
f_n^B\left(z\right)&=&N_n^B\cos\left[B_n\left(z-L/2\right)\right]\\
f^C\left(z\right)&=&N^C\cosh\left[C\left(z-L/2\right)\right]
\end{eqnarray}
where the wavevectors, for $n>0$, are:
\begin{equation}
A_n=\frac{2\pi}{L}n ,
\end{equation}
$B_n$ is implicitly defined by:
\begin{equation}
\label{eq_def_Bn}
-\frac{1}{2}h\frac{\rho_0}{\rho_1}
\frac{\left(1-\frac{\nu_0}{\nu_1}\right)q^2+B_n^2}{B_n}
=
\tan\left(B_n\frac{L}{2}\right) ,
\end{equation}
and the only value of $C$ is implicitly defined by:
\begin{equation}
\label{eq_def_Cn}
\frac{1}{2}h\frac{\rho_0}{\rho_1}
\frac{\left(1-\frac{\nu_0}{\nu_1}\right)q^2-C^2}{C}
=
\tanh\left(C\frac{L}{2}\right)
\end{equation}
The normalization constants are:
\begin{eqnarray}
N_n^A&=&1/\sqrt{\frac{L\rho_1}{2h\rho_0}}\\
N_n^B&=&
1/\sqrt{\frac{L\rho_1}{2h\rho_0}+\frac{1}{2}\cos^2
\left(B_n\frac{L}{2}\right)
\frac{\displaystyle B_n^2-\left(1-\frac{\nu_0}{\nu_1}\right)q^2}{\displaystyle B_n^2}}\\
N^C&=&
1/\sqrt{\frac{L\rho_1}{2h\rho_0}+\frac{1}{2}\cosh^2
\left(C\frac{L}{2}\right)
\frac{\displaystyle C^2+\left(1-\frac{\nu_0}{\nu_1}\right)q^2}{\displaystyle C^2}
}
\end{eqnarray}
The eigenvalues are:
\begin{eqnarray}
\lambda_n^A&=&-\nu_1\left(q^2+A_n^2\right)\\
\lambda_n^B&=&-\nu_1\left(q^2+B_n^2\right)\\
\lambda^C&=&-\nu_1\left(q^2-C^2\right)
\end{eqnarray}
The eigenfunctions are ortogonal with respect to the scalar 
product:
\begin{equation}
\label{eq_def_prodotto_scalare}
f\cdot g=
\frac{\rho_1}{h\rho_0}
\int{f\left(z\right)g\left(z\right)dz}
+f\left(0\right)g\left(0\right)
\end{equation}

The evolution equations for the coefficients are:
\begin{equation}
\label{eq_evoluzione_a_b}
\left\{
\begin{array}{l}
\frac{\partial}{\partial t}a_n^{\vec{q}}\left(t\right)=
\lambda_n^Aa_n^{\vec{q}}\left(t\right)\\
\frac{\partial}{\partial t}b_n^{\vec{q}}\left(t\right)=
\lambda_n^Bb_n^{\vec{q}}\left(t\right)\\
\frac{\partial}{\partial t}c^{\vec{q}}\left(t\right)=
\lambda^Cb_n^{\vec{q}}\left(t\right)
\end{array}
\right.
\end{equation}

In order to use equipartition theorem, I evaluate the kinetic energy
of the fluid:
\begin{equation}
E=
\frac{1}{2}\rho_0h\int{\left|\vec{u}_0\left(\vec{x},t\right)
\right|^2d\vec{x}}+
\frac{1}{2}\rho_1\int{\left|\vec{u}_1\left(\vec{x},z,t\right)
\right|^2d\vec{x}dz}
\end{equation}
In Fourier space, the energy can be decomposed as the sum 
$E_T+E_Z$, where:
\begin{equation}
E_T=
\frac{\left(2\pi\right)^2}{2}\rho_0h
\int{\left|u_0\left(\vec{q},t\right)\right|^2d\vec{q}}+
\frac{\left(2\pi\right)^2}{2}\rho_1
\int{\left|u_1\left(\vec{q},z,t\right)\right|^2d\vec{q}dz}
\end{equation}
The integrals can be expressed in terms of the scalar product
defined in Eq. (\ref{eq_def_prodotto_scalare}):
\begin{equation}
E_T=
\frac{\left(2\pi\right)^2}{2}\rho_0h
\int{u\left(\vec{q},z,t\right)\cdot 
u\left(\vec{q},z,t\right)d\vec{q}}
\end{equation}
Due to the orthonormality of the eigenfunctions I used:
\begin{equation}
E_T=
\frac{\left(2\pi\right)^2}{2}\rho_0h
\int{\left\{\sum_{n=1}^{+\infty}
{\left[
\left|a_n^{\vec{q}}\right|^2+\left|b_n^{\vec{q}}\right|^2
\right]+
\left|c^{\vec{q}}\right|^2
}
\right\}d\vec{q}}
\end{equation}
From equipartition theorem:
\begin{eqnarray}
\left<b_n^{\vec{q}}b_m^{\vec{q}'}\right>
&=&\delta_{n,m}
\delta\left(\vec{q}+\vec{q}'\right)
\frac{K_BT}{\left(2\pi\right)^2\rho_0h}\\
\left<c^{\vec{q}}c^{\vec{q}'}\right>
&=&
\delta\left(\vec{q}+\vec{q}'\right)
\frac{K_BT}{\left(2\pi\right)^2\rho_0h}
\end{eqnarray}

From Eq. (\ref{eq_evoluzione_a_b}) I can derive the time spectrum:
\begin{eqnarray}
\left<b_n^{\vec{q}}\left(\omega\right)b_m^{\vec{q}'}\left(\omega'\right)\right>
&=&\delta_{n,m}
\delta\left(\vec{q}+\vec{q}'\right)
\delta\left(\omega+\omega'\right)
\frac{K_BT}{4\pi^3\rho_0h}
\frac{-\lambda_n^B}{\omega^2+\lambda_n^{B2}}\\
\left<c^{\vec{q}}\left(\omega\right)c^{\vec{q}'}\left(\omega'\right)\right>
&=&
\delta\left(\vec{q}+\vec{q}'\right)
\delta\left(\omega+\omega'\right)
\frac{K_BT}{4\pi^3\rho_0h}
\frac{-\lambda^C}{\omega^2+\lambda^{C2}}
\end{eqnarray}
Using Eq. (\ref{eq_scomposizione_u0}), I evaluate the power
dynamic spectrum of the velocity fluctuations of $\vec{u}_0$:
\begin{eqnarray}
S_u\left(\vec{q},\omega\right)
&=&
\frac{K_BT}{4\pi^3\rho_0h}
\sum_{n=1}^{+\infty}
{\frac{-\lambda_n^B}{\omega^2+\lambda_n^{B2}}
N_n^{B2} \cos^2\left(B_n\frac{L}{2}\right)}+
\\
&&
\frac{K_BT}{4\pi^3\rho_0h}
\frac{-\lambda^C}{\omega^2+\lambda^{C2}}
N^{C2} \cosh^2\left(C\frac{L}{2}\right)
\end{eqnarray}
where the dynamic power spectrum is defined as follows:
\begin{equation}
\left<u_0\left(\vec{q},\omega\right)
u_0\left(\vec{q}',\omega'\right)\right>
=
\delta\left(\vec{q}+\vec{q}'\right)
\delta\left(\omega+\omega'\right)
S_u\left(\vec{q},\omega\right)
\end{equation}

Now I use Eq. (\ref{eq_def_Bn}) in order to express any trigonometric
function of $B_n$ and $C$:
\begin{eqnarray}
S_u\left(\vec{q},\omega\right)
=\\
\frac{K_BT}{4\pi^3\rho_0h}
\sum_{n=1}^{+\infty}
{\frac{2}{L}
\frac{\nu_1\left(q^2+B_n^2\right)}{\omega^2+\nu_1^2\left(q^2+B_n^2\right)^2}
\frac{B_n^2}{
\frac{\rho_1}{h\rho_0}B_n^2+\frac{1}{4}\frac{h\rho_0}{\rho_1}
\left[B_n^2+\left(1-\frac{\nu_0}{\nu_1}\right)q^2\right]^2
+\frac{1}{L}
\left[B_n^2-\left(1-\frac{\nu_0}{\nu_1}\right)q^2\right]
}}
+\\
\frac{K_BT}{4\pi^3\rho_0h}
2
\frac{\nu_1\left(q^2-C^2\right)}{\omega^2+\nu_1^2\left(q^2-C^2\right)^2}
\frac{C^2}{
L\left\{
\frac{\rho_1}{h\rho_0}C^2-\frac{1}{4}\frac{h\rho_0}{\rho_1}
\left[C^2-\left(1-\frac{\nu_0}{\nu_1}\right)q^2\right]^2
\right\}
+C^2+\left(1-\frac{\nu_0}{\nu_1}\right)q^2
}
\end{eqnarray}
From Eq. (\ref{eq_def_Bn}):
\begin{eqnarray}
B_{n+1}-B_n&=&\frac{2\pi}{L}+o\left(\frac{1}{L}\right)\\
C &=& \sqrt{\left(\frac{\rho_1}{h\rho_0}\right)^2+\left(1-\frac{\nu_0}{\nu_1}\right)q^2} 
-\frac{\rho_1}{h\rho_0}+o\left(\frac{1}{L}\right)
\end{eqnarray}
For $L\to \infty$:
\begin{eqnarray}
S_u\left(\vec{q},\omega\right)
=\\
\frac{K_BT}{4\pi^3\rho_0h}
\frac{1}{\pi}
\int_0^{+\infty}
{
\frac{\nu_1\left(q^2+B^2\right)}{\omega^2+\nu_1^2\left(q^2+B^2\right)^2}
\frac{B^2}{
\frac{\rho_1}{h\rho_0}B^2+\frac{1}{4}\frac{h\rho_0}{\rho_1}
\left[\left(1-\frac{\nu_0}{\nu_1}\right)q^2+B^2\right]^2
}dB}+\\
\frac{K_BT}{4\pi^3\rho_0h}
2
\frac{\nu_1\left(q^2-C^2\right)}{\omega^2+\nu_1^2\left(q^2-C^2\right)^2}
\frac{C^2}{
C^2+\left(1-\frac{\nu_0}{\nu_1}\right)q^2
}
\end{eqnarray}
I define:
\begin{eqnarray}
\alpha&=&1-\frac{\nu_0}{\nu_1}
\\
q_0&=&\frac{\rho_1}{h\rho_0}
\\
h\left(x,y\right)&=&\frac{4}{\pi}\int_0^{+\infty}
\frac{1}{\pi}
{\frac{x^2+B^2}{y^2+\left(x^2+B^2\right)^2}
\frac{B^2}{4B^2+\left(\alpha x^2+B^2\right)^2}dB}
\\
\Omega\left(x\right)&=&x^2+2\frac{1}{1-\alpha}
\left(\sqrt{1+\alpha x^2}-1\right)
\\
l\left(x,y\right)&=&\frac{1}{\pi}
\frac{\Omega\left(x\right)}{y^2+\Omega^2\left(x\right)}
\\
A\left(x\right)&=&1-\frac{1}{\sqrt{1+\alpha x^2}}
\end{eqnarray}
and obtain:
\begin{equation}
S_u\left(\vec{q},\omega\right)
=
\frac{K_BT}{4\pi^2\rho_0h}
\left[
\frac{1}{\nu_1q_0^2}h\left(\frac{q}{q_0},
\frac{\omega}{\nu_1q_0^2}\right)
+
\frac{1}{\nu_0q_0^2}
A\left(\frac{q}{q_0}\right)
l\left(\frac{q}{q_0},
\frac{\omega}{\nu_0q_0^2}\right)
\right]
\label{eq_spettro_q_omega}
\end{equation}
The term with $l\left(x,y\right)$ has a Lorentzian behaviour, and
represents the spontaneous velocity fluctuations of the film, damped by the
surrounding fluid. The other one describes the velocity fluctuations
induced by the fluid on the film. It's worth noting that the only spatial
lengthscale is $1/q_0$, while two different temporal lengthscales are involved 
in the two processes: $\nu_0q_0^2$ for the fluctuations of the film,
and $\nu_1q_0^2$ for the velocity fluctuations of the surrounding fluid.

Some explicit results can be obtained by evaluating the integral:
\begin{equation}
\frac{4}{\pi}\int_0^{+\infty}
{\frac{1}{x^2+\alpha B^2}
\frac{B^2}{4B^2+\left(x^2+B^2\right)^2}dB}
=
\frac{1}{x\sqrt{x^2+1}}
\frac{1}{2\sqrt{\alpha}\sqrt{x^2+1}+\left(\alpha+1\right)x}.
\label{eq_integrale_risolto}
\end{equation}
With $\alpha=0$, I obtain, as expected from the equipartition theorem:
\begin{equation}
\int_{-\infty}^{+\infty}
{S_u\left(\vec{q},\omega\right)d\omega}
=
\frac{K_BT}{4\pi^2\rho_0h}
\end{equation}

The power spectrum on $\omega=0$ can be evaluated by 
means of Eq. (\ref{eq_integrale_risolto}):
\begin{equation}
S_u\left(\vec{q},\omega=0\right)
=
\frac{K_BT}{4\pi^3\eta_0h}
\frac{1}{q\left(q_C+q\right)} ,
\label{eq_spettro_statico}
\end{equation}
where:
\begin{equation}
q_C=\frac{2\eta_1}{h\eta_0},
\end{equation}
$K_B$ is the Boltzmann constant, $T$ the temperature, $eta_0$ and
$eta_1$ are the shear viscosity of the film and the surrounding fluid,
and $h$ is the thickness of the film. 

The power spectrum has a $q^{-2}$ dependence for $q\gg q_C$; at smaller
wavevectors, the divergence saturates to $q^{-1}$, due to air damping.
For a soap film in air, the values of densities are $\rho_0=10^3
\mathrm{Kg}/\mathrm{m}^3$
and $\rho_1=1.3\mathrm{Kg}/\mathrm{m}^3$. The
effective viscosity of the film is $\nu_0=1.6\cdot10^{-6}
\mathrm{m}^2/\mathrm{s}$,
\cite{rivera00}, evaluated by means of Trapeznikov relation
\cite{trapeznikov57}. The viscosity of air is
$\nu_1=1.43\cdot10^{-5}\mathrm{m}^2/\mathrm{s}$ \cite{handbook_chem_phys}.
From these values, for a $2\mu\mathrm{m}$ thick film, 
$q_C\approx 1.2\cdot10^4\mathrm{m}^{-1}$, corresponding to a 
$0.5\mathrm{mm}$ wavelength.

From the power spectrum of the velocity fluctuations, we can derive
the diffusion coefficien\cite{brogioli01}:
\begin{equation}
D \approx \pi \int_{\Lambda}{S_u\left(q,\omega=0\right)d^2q}
\end{equation}
where the cut off wavevector $\Lambda$ is of the order of the inverse of
$a$, the radius of the diffusing particle. By integrating the 
static power spectrum of Eq. (\ref{eq_spettro_statico}):
\begin{equation}
D=
\frac{K_BT}{2\pi\eta_0 h}
\left[\ln\frac{h\eta_0}{a\eta_1} - \ln\left(2\right)\right]
\end{equation}
This is the well known Saffman formula for the diffusion coefficient
of a particle on a liquid film \cite{saffman76,cheung96,bechhoefer97}.

This result can be compared with a phenomenological theory \cite{rivera00}.

\section{Time correlation function}

By Fourier anti-transforming Eq.(\ref{eq_spettro_q_omega}) in
$\omega$, I obtain the time velocity time corrlation function:
\begin{equation}
S_u\left(\vec{q},\tau\right)
=
\frac{K_BT}{4\pi^2\rho_0h}
\left[
e^{-\nu_1q^2\left|\tau\right|}
h\left(\alpha\nu_1q_0^2\left|\tau\right|\right)
+A
e^{-\nu\left(x\right)q^2\left|\tau\right|}
\right]
\end{equation}
where:
\begin{eqnarray}
\alpha&=&1-\frac{\nu_0}{\nu_1}
\\
h\left(y\right)&=&\frac{4}{\pi}\int_0^{+\infty}{
e^{-yB^2}
\frac{B^2}{4B^2+\left(x^2+B^2\right)^2}dB}
\\
\nu\left(x\right)&=&
\nu_0
+
\left(\nu_1-\nu_0\right) 2\frac{\sqrt{1+x^2}-1}{x^2}
\\
x&=&\frac{q}{q_0}
\\
A\left(x\right)&=&1-\frac{1}{\sqrt{1+x^2}}
\\
q_0&=&\frac{\rho_1}{h\rho_0}
\frac{1}{\sqrt{\alpha}}
\end{eqnarray}

\bibliography{film.bib}

\end{document}